\documentclass[
    aps,
    prl,
    reprint,
    superscriptaddress
]{revtex4-2}

\usepackage{graphicx}
\usepackage{bm}
\usepackage{dcolumn}
\usepackage{amssymb}
\usepackage{amsmath}
\usepackage{parskip}
\usepackage{array,multirow}
\usepackage{multirow}
\usepackage{dcolumn}    
\usepackage{afterpage}
\usepackage{rotating}
\usepackage[hidelinks]{hyperref}
\usepackage{xcolor}
\usepackage{titlesec}
\usepackage{float}
\usepackage{ifxetex,ifluatex}
\usepackage{xcolor}
\usepackage[english]{babel}
\usepackage[sort&compress]{natbib}
\usepackage{xfrac}
\usepackage{graphicx}
\usepackage{threeparttable}
\usepackage{mathpazo}
\usepackage{textcomp}
\usepackage{nicefrac} 
\usepackage[perpage]{footmisc} 
\usepackage{pifont}
\newcommand{\cmark}{\ding{51}}
\newcommand{\xmark}{\ding{55}}
\begin{document}

\title{Light-induced chiral altermagnets from conventional antiferromagnets
}

\author{Jorge Cardenas-Gamboa}
\email{jicg1@ifw-dresden.de}
\affiliation{Leibniz Institute for Solid State and Materials Research Dresden, 01069 Dresden, Germany}

\author{Devika Jayadeep}
\affiliation{D\'epartement de physique et Institut quantique, Universit\'e de Sherbrooke, Sherbrooke J1K 2R1 QC, Canada}

\author{Aritz Leonardo}
\affiliation{ Donostia International Physics Center, Donostia-San Sebastian 20018 Gipuzkoa, Spain}
\affiliation{EHU Quantum Center, University of the Basque Country UPV/EHU, 48940 Leioa, Spain}

\author{Gregory A. Fiete}
\affiliation{Department of Physics, Northeastern University, Boston, Massachusetts 02115, USA}
\affiliation{Quantum Materials and Sensing Institute, Northeastern University, Burlington, Massachusetts 01803, USA}
\affiliation{Department of Physics, Massachusetts Institute of Technology, Cambridge, MA 02139, USA}
\affiliation{Department of Physics, Harvard University, Cambridge, MA 02138, USA}

\author{Jeroen van den Brink}
\affiliation{Leibniz Institute for Solid State and Materials Research Dresden, 01069 Dresden, Germany}
\affiliation{Würzburg-Dresden Cluster of Excellence Ct.qmat, Technische Universitat Dresden, 01062, Dresden, Germany}

\author{Maia G. Vergniory}
\email{maia.vergniory@usherbrooke.ca}
\affiliation{D\'epartement de physique et Institut quantique, Universit\'e de Sherbrooke, Sherbrooke J1K 2R1 QC, Canada}
\affiliation{ Donostia International Physics Center, Donostia-San Sebastian 20018 Gipuzkoa, Spain}
\affiliation{Regroupement Qu\'eb\'ecois sur les Mat\'eriaux de Pointe (RQMP), Quebec H3T 3J7, Canada}

\date{\today}

\begin{abstract}

Altermagnets combine compensated collinear magnetic order with momentum-dependent nonrelativistic spin splitting, offering new opportunities for spintronic applications. Here, we propose a route to optically induce altermagnetism through coherent lattice control. Using symmetry analysis, we identify phonon distortions that lift the spin degeneracy of conventional antiferromagnets while preserving the rotational symmetries connecting opposite-spin sublattices. Among these, geometric chiral phonons provide a particularly appealing pathway, linking the emergence of altermagnetism to structural chirality. Taking Fe$_2$TeO$_6$ as a prototype, we demonstrate that a linearly polarized mid-infrared pulse can drive a structural instability through negative biquadratic coupling between an infrared-active phonon and an optically silent, symmetry-breaking mode. First-principles-based lattice dynamics reveal the emergence of a transient chiral distortion, while electronic-structure calculations establish that the light-induced structure exhibits nonrelativistic spin splitting characteristic of altermagnetism. Our results establish a symmetry-guided pathway toward the ultrafast optical control of altermagnetic states through nonlinear phononics.

\end{abstract}

\maketitle

\vspace{-10pt}
\section{Introduction}
\vspace{-10pt}

Altermagnets have attracted broad attention because they combine spin-split electronic bands with compensated collinear magnetic order~\cite{smejkal_emerging,smejkal_beyond,mazin_editorial,mazin_fesb2}. Unlike conventional antiferromagnets, their opposite-spin sublattices are connected by crystallographic rotations rather than by inversion or translation. The resulting spin splitting alternates in sign across the Brillouin zone and persists in the absence of spin-orbit coupling~\cite{hayami_jpsj,yuan_prb,yuan_prm,naka_organic}. Altermagnets can support spin-polarized currents and spin-splitter torques~\cite{gonzalez_splitter,bai_torque}, as well as an anomalous Hall effect when allowed by magnetic symmetry~\cite{smejkal_crystal}. These transport properties, together with their compensated magnetization and terahertz spin dynamics, make altermagnets attractive for spintronics~\cite{alter_spitronic,hu2025spin,song2025altermagnets}. This raises a central question: can a conventional, spin-degenerate antiferromagnet be transformed into an altermagnet dynamically, while preserving its compensated magnetic order?

Altermagnetic spin splitting depends on both crystal structure and magnetic order, offering a route to control through lattice distortions. Advances in optical lattice control raise the prospect of reversibly inducing this splitting in an initially spin-degenerate antiferromagnet, complementing static structural tuning by strain, pressure, or chemical substitution~\cite{antiferroaxial_alter}. Intense mid-infrared and terahertz pulses can resonantly drive infrared-active phonons to large amplitudes, coherently reshaping a crystal on ultrafast timescales~\cite{disa_review,non_linear_1,subedi_lightcontrol}. Through nonlinear phononics, the driven mode couples anharmonically to other lattice coordinates, generating transient structural distortions~\cite{non_linear_2,MnPS3_nonlinear}. Such mode-selective lattice control offers a route to dynamically induce altermagnetism by lifting spin degeneracy while preserving compensated collinear magnetic order.

\begin{table*}[t]
\centering
\renewcommand{\arraystretch}{1.4}
   \caption{Symmetry-lowering pathways from centrosymmetric parent point groups $G$ to non-centrosymmetric subgroups $H$ through the condensation of odd-parity phonon modes $Q_{\text{AM}}$. In all cases, the parent antiferromagnetic phase is considered to possess a $\mathcal{PT}$-type protection associated with the spin-group operation $[C_2\Vert P]$. The resulting subgroups remove inversion symmetry while preserving operations that can connect the opposite-spin sublattices, as required for altermagnetism. The GCP column indicates whether the condensed mode is a geometric chiral phonon.}
\label{tab:afm_fm}
\begin{tabular}{ccccc}
\hline
\textbf{Parent Point Group ($G$)} &
\textbf{Altermagnetic Mode ($Q_{\rm AM}$)} &
\textbf{Final Point Group ($H$)} &
\textbf{GCP} &
\textbf{Example Materials} \\
\hline

$C_{2h}(2/m)$ &
\begin{tabular}[c]{@{}c@{}}
$A_u$\\
$B_u$
\end{tabular} &
\begin{tabular}[c]{@{}c@{}}
$C_2(2)$\\
$C_s(m)$
\end{tabular} &
\begin{tabular}[c]{@{}c@{}}
\cmark\\
\xmark
\end{tabular} &
FeWO$_4$
\\
\hline

$D_{2h}(mmm)$ &
\begin{tabular}[c]{@{}c@{}}
$A_u$\\
$B_{1u},\,B_{2u},\,B_{3u}$
\end{tabular} &
\begin{tabular}[c]{@{}c@{}}
$D_2(222)$\\
$C_{2v}(mm2)$
\end{tabular} &
\begin{tabular}[c]{@{}c@{}}
\cmark\\
\xmark
\end{tabular} &
LiMnPO$_4$
\\
\hline

$C_{4h}(4/m)$ &
\begin{tabular}[c]{@{}c@{}}
$A_u$\\
$B_u$
\end{tabular} &
\begin{tabular}[c]{@{}c@{}}
$C_4(4)$\\
$S_4(\bar{4})$
\end{tabular} &
\begin{tabular}[c]{@{}c@{}}
\cmark\\
\xmark
\end{tabular} &
KOsO$_4$, KRuO$_4$
\\
\hline

$D_{4h}(4/mmm)$ &
\begin{tabular}[c]{@{}c@{}}
$A_{1u}$\\
$B_{1u},\,B_{2u}$\\
$A_{2u}$
\end{tabular} &
\begin{tabular}[c]{@{}c@{}}
$D_4(422)$\\
$D_{2d}(\bar{4}2m)$\\
$C_{4v}(4mm)$
\end{tabular} &
\begin{tabular}[c]{@{}c@{}}
\cmark\\
\xmark\\
\xmark
\end{tabular} &
Bi$_2$CuO$_4$, LiMnAs, KMnF$_3$
\\
\hline

$D_{3d}(\bar{3}m)$ &
\begin{tabular}[c]{@{}c@{}}
$A_{1u}$\\
$A_{2u}$
\end{tabular} &
\begin{tabular}[c]{@{}c@{}}
$D_3(32)$\\
$C_{3v}(3m)$
\end{tabular} &
\begin{tabular}[c]{@{}c@{}}
\cmark\\
\xmark
\end{tabular} &
Cr$_2$O$_3$, SrRu$_2$O$_6$
\\
\hline

$C_{6h}(6/m)$ &
\begin{tabular}[c]{@{}c@{}}
$A_u$\\
$E_{2u}$\\
$B_u$
\end{tabular} &
\begin{tabular}[c]{@{}c@{}}
$C_6(6)$\\
$C_2(2)$\\
$C_{3h}(\bar{6})$
\end{tabular} &
\begin{tabular}[c]{@{}c@{}}
\cmark\\
\cmark\\
\xmark
\end{tabular} &
PbMn$_2$Ni$_6$Te$_3$O$_{18}$
\\
\hline

$D_{6h}(6/mmm)$ &
\begin{tabular}[c]{@{}c@{}}
$A_{1u}$\\
$B_{2u}$\\
$A_{2u}$
\end{tabular} &
\begin{tabular}[c]{@{}c@{}}
$D_6(622)$\\
$D_{3h}(\bar{6}m2)$\\
$C_{6v}(6mm)$
\end{tabular} &
\begin{tabular}[c]{@{}c@{}}
\cmark\\
\xmark\\
\xmark
\end{tabular} &
ScMn$_6$Ge$_6$
\\
\hline

$O_h(m\bar{3}m)$ &
\begin{tabular}[c]{@{}c@{}}
$A_{1u}$\\
$A_{2u}$
\end{tabular} &
\begin{tabular}[c]{@{}c@{}}
$O(432)$\\
$T_d(\bar{4}3m)$
\end{tabular} &
\begin{tabular}[c]{@{}c@{}}
\cmark\\
\xmark
\end{tabular} &
TmGa$_{3}$
\\
\hline

\end{tabular}
\end{table*}

In this work, we combine group-subgroup and spin-symmetry analyses with first-principles modeling of driven lattice dynamics to propose phonon-mediated routes from antiferromagnetism to altermagnetism. We identify odd-parity phonon distortions that lower centrosymmetric parent structures to noncentrosymmetric subgroups while satisfying the spin-symmetry requirements for altermagnetism, and we tabulate representative pathways (Table~\ref{tab:afm_fm}). Among these, geometric chiral phonons~\cite{romao2024phonon} are particularly appealing because their distortions also induce structural chirality, yielding two enantiomeric altermagnetic structures. We demonstrate access to such a distortion in Fe$_2$TeO$_6$ through biquadratic nonlinear phonon coupling~\cite{biquadra_1,biquadra_2}. In particular we show how an optically driven infrared mode can soften the geometric chiral coordinate and induce a transient structural instability above a driving threshold. Using parameters obtained from density-functional theory, we simulate the lattice response to a mid-infrared pulse and calculate the electronic bands of the distorted structure, revealing the light-induced momentum-dependent nonrelativistic spin splitting characteristic of altermagnetism.

\begin{figure*}[t]
    \centering
    \includegraphics[width=\linewidth]{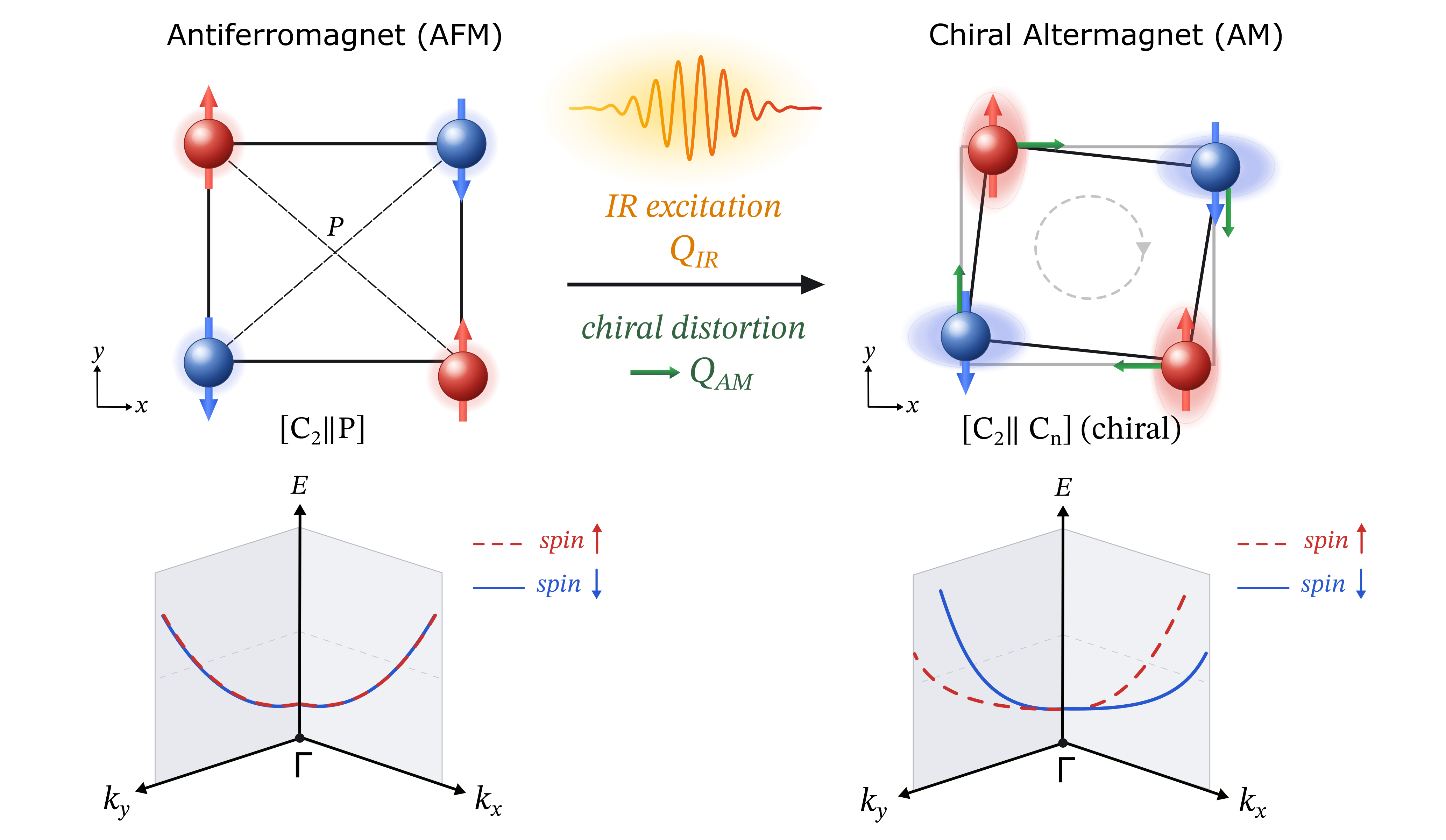}
    \caption{Schematic illustration of the symmetry-guided route from a conventional antiferromagnet (AFM) to an altermagnetic (AM) state. The upper panels show the symmetry reduction produced by the condensation of a symmetry-breaking altermagnetic mode ($Q_{\mathrm{AM}}$), which removes the inversion-based ($[C_2\Vert P]$) protection of the AFM phase while preserving operations relating opposite-spin sublattices. The lower panels show the corresponding evolution from spin-degenerate bands in the AFM phase to momentum-dependent spin-split bands in the AM phase.}
    \label{fig:mechanism}
\end{figure*}

\section{Results and discussion}
\noindent $Symmetry~Framework$: Conventional collinear antiferromagnets and altermagnets differ in the symmetries that relate their opposite-spin sublattices. In the nonrelativistic limit, spin-group operations combining a $180^\circ$ spin rotation with spatial inversion or a sublattice-exchanging translation, $[C_2\Vert P]$ and $[C_2\Vert t]$, protect spin degeneracy throughout the Brillouin zone~\cite{smejkal_beyond,smejkal_emerging}. Here, $C_2$ denotes a spin rotation about an axis perpendicular to the magnetic moments, while $P$ and $t$ denote spatial inversion and a translation, respectively. Altermagnets lack these degeneracy-protecting operations but retain symmetries combining spin reversal with proper or improper crystallographic rotations that connect opposite-spin sublattices. This allows momentum-dependent spin splitting while preserving exactly compensated magnetic order.

The transition from a conventional antiferromagnet to an altermagnet can therefore be described in terms of spin-group subgroup reductions~\cite{afm_am_zhou}. The essential requirement is to remove the symmetries that enforce spin degeneracy throughout the Brillouin zone while retaining rotational operations connecting the opposite-spin sublattices. In the lattice-driven pathways considered here, a phonon distortion lowers the crystallographic point group from $G$ to a subgroup $H$. Importantly, this crystallographic symmetry reduction alone is not sufficient to establish altermagnetism: the spin group of the distorted magnetic structure must also satisfy the corresponding symmetry requirements.

The next question is how to realize these symmetry reductions in a controlled manner. Previous proposals have considered static electric fields~\cite{electr_2}, Floquet engineering~\cite{floquet_alter,floquet_theory2}, and surface engineering~\cite{surface_eng_1,surface_eng_2}. Here, we instead exploit coherent lattice dynamics. A phonon distortion selects an isotropy subgroup of the parent crystal symmetry, providing a systematic route to the structural symmetry reductions required for an antiferromagnet-to-altermagnet transition. We refer to phonon modes capable of enabling such a transition as altermagnetic modes ($Q_{\mathrm{AM}}$). Their distortions must remove the inversion- and translation-related symmetries protecting spin degeneracy while preserving a proper or improper rotation connecting opposite-spin sublattices. We consider distortions from centrosymmetric parent structures to both chiral and achiral noncentrosymmetric subgroups, with particular emphasis on the chiral pathways, which remove all improper crystallographic operations. For a fixed collinear magnetic order without $t\mathcal{T}$ symmetry, the latter pathways can enable altermagnetism provided that the surviving proper rotations connect opposite-spin sublattices and preserve zero net magnetization. Table~\ref{tab:afm_fm} summarizes representative phonon-mediated symmetry-lowering pathways.

\begin{figure*}[t]
    \centering
    \includegraphics[width=\linewidth]{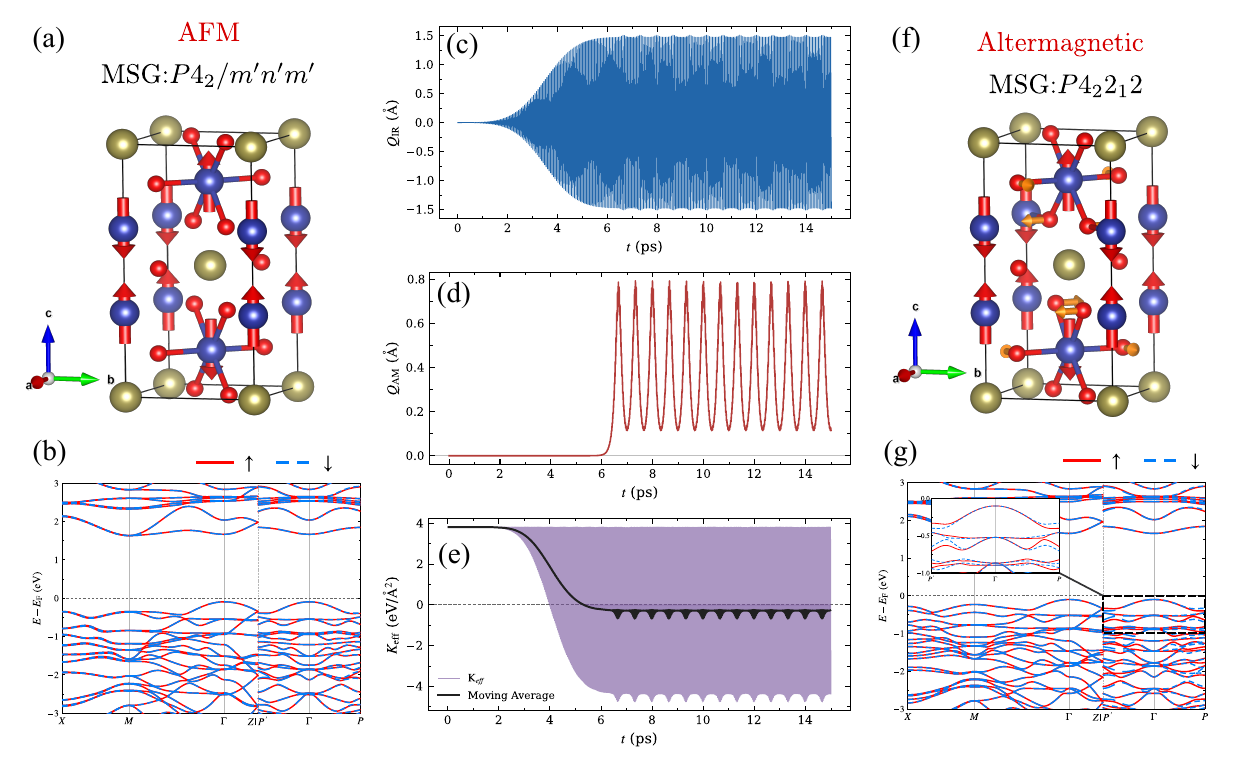}
    \caption{ (a,b) Parent antiferromagnetic structure and spin-resolved band structure. Blue, gold, and red spheres denote Fe, Te, and O, respectively. Red arrows indicate magnetic moments. (c--e) Time evolution of the infrared coordinate $Q_{\mathrm{IR}}$ ($E_u$), altermagnetic coordinate $Q_{\mathrm{AM}}$ ($A_{1u}$), and effective force constant $K_{\mathrm{eff}}$ under a linearly polarized Gaussian pulse with $E_0=7.4$~MV/cm and $\tau_{\mathrm{FWHM}}=1.0$~ps. Purple and black curves in (e) show the effective force constant and its moving average, respectively. (f,g) $A_{1u}$-distorted altermagnetic structure and corresponding band structure along the same momentum path as in (b). Orange arrows indicate atomic displacements. The inset in (g) provides an enlarged view of the region showing spin splitting.}
    \label{fig:nonlinearphononics}
\end{figure*}

Of particular interest are geometric chiral phonons (GCPs), whose nonlinear excitation has been shown theoretically to induce transient structural chirality~\cite{romao2024phonon}. In contrast to phonons characterized by circular atomic motion, their chirality is encoded in the displaced atomic geometry itself. For a centrosymmetric parent, opposite signs of an odd-parity GCP coordinate produce enantiomeric structures belonging to one of the 11 chiral point groups. Table~\ref{tab:afm_fm} distinguishes modes leading to chiral structures (\cmark) from those producing achiral structures (\xmark). Both classes offer candidate routes to dynamically induced altermagnetism, while GCPs additionally link the symmetry reduction to structural chirality.

Figure~\ref{fig:mechanism} illustrates the proposed symmetry pathway. In the parent antiferromagnet, the inversion-based spin-group operation $[C_2\Vert P]$ relates opposite-spin sublattices and protects spin degeneracy throughout the Brillouin zone. A distortion along $Q_{\mathrm{AM}}$ removes this protection while retaining crystallographic rotations that connect opposite-spin sublattices. The resulting symmetry reduction permits spin splitting at generic crystal momenta without introducing a net magnetization. We now turn to the nonlinear-phononics mechanism by which an optically excited infrared-active mode can drive this symmetry-breaking distortion.

\noindent $Nonlinear~Phononics~Framework$: In conventional nonlinear phononics, a coherently excited infrared-active phonon $Q_{\mathrm{IR}}$ couples anharmonically to a Raman mode $Q_{\mathrm{R}}$ through a cubic interaction $Q_{\mathrm{IR}}^{2}Q_{\mathrm{R}}$~\cite{non_linear_1,non_linear_2}. This interaction generates a rectified force on the Raman coordinate and has been extensively employed to induce transient structural distortions~\cite{cap1,cap2,cap3}. However, the analogous coupling $Q_{\mathrm{IR}}^{2}Q_{\mathrm{AM}}$ is forbidden by inversion symmetry for the odd-parity altermagnetic modes considered here.


The symmetry-allowed biquadratic interaction $Q_{\mathrm{IR}}^{2}Q_{\mathrm{AM}}^{2}$ allows the driven infrared mode to modulate the effective stiffness of the altermagnetic coordinate~\cite{biquadra_1}. For a coupling that softens this coordinate, sufficiently strong excitation can destabilize the parent structure and induce a transient symmetry-breaking distortion along $Q_{\mathrm{AM}}$~\cite{biquadra_2}.

The dynamics of the driven lattice may be described using the minimal potential

\begin{equation}
\begin{split}
    V(Q_{\mathrm{IR}},Q_{\mathrm{AM}};t)
    &= \frac{1}{2}K_{\mathrm{IR}}Q_{\mathrm{IR}}^2
    + \frac{1}{2}K_{\mathrm{AM}}Q_{\mathrm{AM}}^2 + \frac{1}{4}\beta Q_{\mathrm{AM}}^4\\
    &\quad 
    + DQ_{\mathrm{IR}}^2Q_{\mathrm{AM}}^2
    - Z^{*}Q_{\mathrm{IR}}E(t),
\end{split}
\label{eq:biquadratic}
\end{equation}

where $K_{\text{IR}}$ and $K_{\text{AM}}$ are the harmonic force constants of the infrared (IR) and symmetry-breaking modes respectively, $\beta$ is the quartic anharmonic coefficient that stabilizes large-amplitude distortions, and $D$ is the biquadratic coupling strength between the two phonon coordinates. The last term is the light-matter interaction, $-\Delta\vec{P}\cdot\vec{E}(t) = -Z^{*}Q_{\text{IR}}E(t)$, where $Z^{*}Q_{\text{IR}}$ is the induced dipole moment of the IR mode, $Z^{*}$ the Born effective charge, and $E(t)$ the electric field component of the Gaussian laser pulse.

The biquadratic coupling modifies the curvature of the potential along $Q_{\mathrm{AM}}$, giving an effective force constant
\begin{equation}
    K_{\mathrm{eff}}(t)=K_{\mathrm{AM}}+2DQ_{\mathrm{IR}}^{2}(t).
    \label{eq:k_eff}
\end{equation}
For $D>0$, displacement of the infrared mode stiffens the altermagnetic mode, whereas $D<0$ softens it. At fixed $Q_{\mathrm{IR}}$, sufficiently strong softening, $2|D|Q_{\mathrm{IR}}^{2}>K_{\mathrm{AM}}$, destabilizes the parent configuration and, for $\beta>0$, produces minima at finite $Q_{\mathrm{AM}}$. Negative biquadratic coupling therefore provides a route to inducing an altermagnetic distortion. Because the infrared motion is oscillatory, we assess whether this distortion develops by solving the coupled lattice dynamics.

\noindent $Driven~lattice~dynamics~in~Fe_2TeO_6$: To explore and illustrate its practical applicability, we apply this framework to Fe$_2$TeO$_6$, a collinear antiferromagnet~\cite{fe2teo6} with magnetic space group $P4_2/m'n'm'$ (No.~136.503) (see Figure~\ref{fig:nonlinearphononics}(a)). Its combined inversion-time-reversal symmetry $\mathcal{PT}$ protects twofold band degeneracy throughout the Brillouin zone, consistent with the coincident spin-up and spin-down bands in our calculations (see Figure~\ref{fig:nonlinearphononics}(b)). The lattice vibration representation of the crystallographic point group $D_{4h}$ is given by the equation

\begin{equation}
\begin{aligned}
\Gamma_{\mathrm{vib}} ={}&
4A_{1g} \oplus 2A_{2g} \oplus 2B_{1g}
\oplus 4B_{2g} \oplus 6E_g \\
&\oplus A_{1u} \oplus 5A_{2u}
\oplus 5B_{1u} \oplus B_{2u} \oplus 9E_u .
\end{aligned}
\end{equation}

This representation includes three acoustic modes, $A_{2u}\oplus E_u$, leaving 51 optical modes. We focus on coupling the infrared-active $E_u$ modes to the silent $A_{1u}$ mode, identified here as $Q_{\mathrm{AM}}$. For the parent magnetic order, a distortion along the $A_{1u}$ eigenvector lowers the magnetic space group to $P4_2 2_1 2$, breaks $\mathcal{PT}$, and permits altermagnetic spin splitting. Biquadratic coupling is symmetry allowed because $A_{1u}\otimes A_{1u}=A_{1g}$ and the symmetric product of $E_u$ with itself contains $A_{1g}$. The corresponding invariant is $Q_{\mathrm{AM}}^2(Q_{\mathrm{IR},x}^2+Q_{\mathrm{IR},y}^2)$. For linearly polarized excitation, we retain the driven component, denoted by $Q_{\mathrm{IR}}$, recovering the two-mode potential in Eq.~(\ref{eq:biquadratic}). We investigate resonant driving of the $E_u$ mode at $f_{\mathrm{IR}}=16.70$~THz, coupled to the $A_{1u}$ mode at $f_{\mathrm{AM}}=7.63$~THz, where $f_\nu=\sqrt{K_\nu/m_\nu}/(2\pi)$. The fitted parameters are summarized in Table~\ref{tab:lattice_parameters}. Details of the first-principles calculations and parameter extraction are provided in the Supplemental Material. 

Figures~\ref{fig:nonlinearphononics}(c--e) show representative lattice dynamics calculated using the DFT-derived parameters under a linearly polarized Gaussian pulse with peak electric field $E_0=7.4$~MV/cm and duration $\tau_{\mathrm{FWHM}}=1.0$~ps. Resonant excitation produces coherent oscillations of the infrared coordinate $Q_{\mathrm{IR}}$ (see Figure~\ref{fig:nonlinearphononics}(c)). Through the negative biquadratic coupling, this motion reduces the effective force constant $K_{\mathrm{eff}}(t)$ (Eq.~(\ref{eq:k_eff})), shown in purple in Figure~\ref{fig:nonlinearphononics}(e). Its moving average, shown in solid black lines, highlights the underlying softening beneath the rapid oscillations. Negative values of $K_{\mathrm{eff}}$ correspond to negative curvature of the potential along $Q_{\mathrm{AM}}$, favoring growth of a symmetry-breaking distortion. The resulting amplification of $Q_{\mathrm{AM}}$ (see Figure~\ref{fig:nonlinearphononics}(d)) demonstrates nonlinear excitation of the $A_{1u}$ mode.

\begin{table}[t]
\caption{
Parameters of the fitted lattice potential and phonon dynamics.
The quartic terms are $\beta Q_{\mathrm{AM}}^4/4$ and
$DQ_{\mathrm{IR}}^2Q_{\mathrm{AM}}^2$.
For the $x$-polarized pulse, $Z^{*}=Z_{\mathrm{IR},x}^{*}$.
All parameters are obtained from DFT calculations or fits to DFT results.
}
\label{tab:lattice_parameters}
\begin{ruledtabular}
\begin{tabular}{lcc}
Parameter & Value & Unit \\
\hline
$K_{\mathrm{AM}}$      & 3.815  & eV/\AA$^{2}$ \\
$K_{\mathrm{IR}}$      & 20.10  & eV/\AA$^{2}$ \\
$\beta$               & 0.857 & eV/\AA$^{4}$ \\
$D$                   & $-1.88$ & eV/\AA$^{4}$ \\
$m_{\mathrm{AM}}$      & 15.99  & amu \\
$m_{\mathrm{IR}}$      & 17.62  & amu \\
$Z_{\mathrm{IR},x}^{*}$ & 3.12   & $e$ \\
\end{tabular}
\end{ruledtabular}
\end{table}

Figure~\ref{fig:nonlinearphononics}(f) illustrates the atomic displacement pattern of the excited $A_{1u}$ mode. The distortion involves in-plane motion of eight oxygen atoms, while the Fe and Te sites remain fixed (see orange arrows in Figure~\ref{fig:nonlinearphononics}(f)). These displacements produce oppositely directed local twists of oxygen pairs surrounding inversion-related Fe sites, modifying their ligand environments and lowering the structural symmetry from $P4_2/mnm$ to the chiral subgroup $P4_2 2_1 2$. For the fixed collinear magnetic order, the distortion removes $\mathcal{PT}$ while retaining proper rotations connecting opposite-spin sublattices, thereby allowing altermagnetic spin splitting with zero net magnetization. The electronic consequence is shown in Fig.~\ref{fig:nonlinearphononics}(g). Along the same momentum path used for the parent antiferromagnet, the $A_{1u}$-distorted structure exhibits nonrelativistic spin splitting, resolved in the enlarged inset which reach values up to $79$~meV. Together, the driven lattice dynamics and the distorted-state band structure demonstrate the proposed route to dynamically inducing altermagnetic spin splitting through nonlinear excitation of the $A_{1u}$ mode.

\vspace{-10pt}

\vspace{-10pt}
\section{Conclusion and outlook}
\vspace{-10pt}

We propose a symmetry-guided route to inducing altermagnetism in a conventional, spin-degenerate antiferromagnet through coherent lattice control. Combining spin-symmetry analysis with first-principles-based lattice dynamics, we identify phonon distortions that remove the symmetries protecting spin degeneracy while retaining the rotational relations between opposite-spin sublattices and, consequently, magnetic compensation. Geometric chiral phonons provide a particularly appealing pathway by linking this transition to structural chirality. In Fe$_2$TeO$_6$, our two-mode simulations show that a linearly polarized mid-infrared pulse, resonantly driving an infrared-active $E_u$ phonon, softens the silent $A_{1u}$ mode through negative biquadratic coupling and produces a transient chiral distortion. The electronic structure calculated for the distorted configuration exhibits the nonrelativistic, momentum-dependent spin splitting characteristic of altermagnetism.

Our results establish a theoretical proof of principle rather than an experimentally demonstrated transition. Although Fe$_2$TeO$_6$ exhibits the required negative biquadratic coupling, the strong driving fields and large-amplitude distortions considered here call for further validation beyond the truncated two-mode potential, including higher-order anharmonicity and coupling to additional lattice modes. Quantitative predictions of the excitation threshold and the transient structural response therefore remain subject to these limitations. Materials closer to a structural instability along an altermagnetic coordinate may offer more favorable conditions, as a reduced restoring force could lower the required driving strength. Van der Waals antiferromagnets with soft interlayer distortions are another promising direction, provided that the relevant displacements fulfill the spin-symmetry requirements and preserve the magnetic order.

An additional opportunity is the control of structural handedness~\cite{induced_chirality,intr_cardenas2025photostriction}. Because the present biquadratic potential is even in $Q_{\mathrm{AM}}$, it does not distinguish between the two enantiomeric distortions. A symmetry-allowed interaction odd in $Q_{\mathrm{AM}}$ could lift this degeneracy and favor one handedness over the other. For a fixed magnetic domain, the parent $\mathcal{PT}$ symmetry relates the two distortions with exchanged spin labels, linking enantiomer selection to reversal of the altermagnetic spin splitting. Such control could ultimately enable optically gated spin-current switches and ultrafast spintronic logic or memory devices~\cite{Mencos2026,gonzalez_splitter,bai_torque}. Realizing these applications will require establishing the lifetime of the induced distortion and the resulting transport response.

\noindent\textbf{Note added.} During the preparation of this manuscript, we became aware of related work by Yuan \textit{et al.}~\cite{Yuan2026}, which investigates the dynamical generation of nonrelativistic spin splitting through nonlinear phononics. In contrast, our work focuses on a transition from conventional antiferromagnetism to altermagnetism mediated by a chiral lattice distortion.

\section{Acknowledgements} 
M.G.V. received financial support from the Canada Excellence Research Chairs Program for Topological Quantum Matter and PID2025-172166NB-C22 funded by MICIU/AEI/10.13039/501100011033. This work has been funded by the Deutsche Forschungsgemeinschaft (DFG, German Research Foundation), through the Wüzburg-Dresden Cluster of Excellence ctd.qmat -- Complexity, Topology and Dynamics in Quantum Matter (EXC 2147, project-id 390858490).
We acknowledge Martin Rodriguez-Vega for discussions.


\bibliography{acs-achemso}

\end{document}